\documentclass[preprint,aps,amsmath,amssymb]{revtex4}

\usepackage{graphicx}
\usepackage{dcolumn}
\usepackage{bm}

\def\laq{\hbox{~}\raise 0.4ex\hbox{$<$}\kern -0.8em\lower 
0.62ex\hbox{$\sim$}\hbox{~}}
\def\gaq{\hbox{~}\raise 0.4ex\hbox{$>$}\kern -0.7em\lower 
0.62ex\hbox{$\sim$}\hbox{~}}
\def\refb#1{(\ref{#1})}
\begin{document}

\title{Black hole multiplicity at particle colliders\\
(Do black holes radiate mainly on the brane?)}
\author{Marco Cavagli\`a}
\altaffiliation{Email: marco.cavaglia@port.ac.uk}
\affiliation{ Institute of Cosmology and Gravitation,  
University of Portsmouth, Portsmouth P01 2EG, United Kingdom}

\begin{abstract}
If gravity becomes strong at the TeV scale, we may have the chance to
produce black holes at particle colliders. In this paper we study some
experimental signatures of black hole production in TeV-gravity theories.
In contrast with the usual lore, we show that the black hole energy loss
in the bulk during the Hawking evaporation phase may be of the same order
of the energy radiated into the brane. We investigate in detail the
multiplicity of the decay products of black hole evaporation. We find that
the number of particles produced during the evaporation phase is
significantly lower than the average multiplicity which has been used in
the past literature.
\end{abstract}

\preprint{Preprint ICG 03/10}

\maketitle

\section{Introduction}

The inauguration of the Large Hadron Collider (LHC) at CERN \cite{LHC} could
well coincide with the grand opening of the first man-made black hole (BH)
factory. Some models of high-energy physics indeed predicts the fundamental
scale of gravity to be as low as a few TeVs \cite{Antoniadis:1990ew,
Arkani-Hamed:1998rs, Antoniadis:1998ig, Randall:1999ee, Randall:1999vf}. If
this is the case, events at energies above this threshold could trigger
nonperturbative gravitational effects, such as the creation of BHs
\cite{Banks:1999gd} and other extended objects predicted by quantum gravity
theories \cite{Ahn:2002mj,Ahn:2002zn}. (For recent reviews, see
\cite{Cavaglia:2002si, Tu:2002xs, Landsberg:2002sa, Emparan:2003xu}.)

Proton-proton collisions at LHC, with center-of-mass energy of $14$ TeV
and luminosity $L=10^{34}$ cm$^{-2}$ s$^{-1}$, could produce BHs at a very
high rate \cite{Giddings:2001bu, Dimopoulos:2001hw, Dimopoulos:2001qe,
Cheung:2002aq, Cheung:2001ue, Park:2001xc, Giudice:2001ce, Casadio:2001wh,
Mocioiu:2003gi}. The detection of these BHs would proceed through the
observation of the decay products of the ensuing Hawking thermal emission
on the brane. The Hawking radiation provides distinct experimental
signatures that would allow to discriminate between gravitational events
and other perturbative nongravitational events \cite{Giddings:2001bu,
Anchordoqui:2002cp, Han:2002yy}.

A number of papers have been devoted to the study of the experimental
signatures of BH formation at particle colliders. The smoking gun of BH
creation would be the detection of events with large multiplicity, high
sphericity, large visible transverse energy and a hadron-to-lepton ratio of
about 5:1 \cite{Giddings:2001bu}, at least in the simplest compactification
models with infinitesimal brane thickness. (For an alternative scenario, see
\cite{Han:2002yy}.) However, an accurate estimate of the physical observables
is often hindered by the poor knowledge of Hawking mechanism and by the use of
a number of crude approximations. Although the solution of the theoretical
conundrums requires the quantum gravity theory, the accuracy of the predictions
can be improved by computational process refinement.

Here we focus on two specific physical observables of BH evaporation: The
multiplicity of the decay products and the BH energy loss. The number of
particles which are produced during the radiation process can be estimated from
dimensional arguments to be of the order of $N\sim M_{\rm BH}/T_{\rm BH}$,
where $M_{\rm BH}$ and $T_{\rm BH}$ are the initial mass and the initial
Hawking temperature of the BH, respectively. The average multiplicity $\langle
N\rangle$ is usually evaluated in the literature by assuming a Boltzmann
statistics for the decay products and neglecting the change in the BH mass
during the radiation process \cite{Dimopoulos:2001hw}. This gives a factor 1/2,
leading to the result $\langle N\rangle=M_{\rm BH}/2T_{\rm BH}$. The Boltzmann
statistics approximation is also assumed in the computation of the BH energy
loss \cite{Emparan:2000rs}. The aim of this paper is to compute the BH energy
loss and the multiplicity to a better precision by dropping the Boltzmann
statistics and the constant-mass assumptions. We also estimate the distribution
of decay products vs.\ the spin, or ``flavor multiplicity'', which is one of
the fundamental observational signatures of high-energy scattering
gravitational events. We find that the bulk energy loss may not be negligible
in some models, and that the total multiplicity may significantly differ from
the average value, depending on the brane dimension. This has important
consequences on the phenomenology of BH creation at particle colliders.  

{\it Notations: Throughout the paper we use natural Planck units with
$G_d=M_{\rm Pl}^{-2}$, where $G_d$ and $M_{\rm Pl}$ are the $d$-dimensional
Newton constant and the Planck mass, respectively.}

\section{Black hole energy loss and multiplicity}

The emission rate for a particle of spin $s_i$ and mass $m_i\ll M$ ``from'' a
BH of mass $M$ into a $n_i$-dimensional slice of the $d$-dimensional spacetime
is described by the blackbody distribution 
\begin{equation}
\frac{d{\cal N}_i}{dt}=\frac{A_i(M,n_i,d) c_i(n_i)\Gamma_i(E,n_i,s_i)}{(2\pi)^{n_i-1}}
\frac{d^{n_i-1}k}{e^{E/T}-(-1)^{2s_i}}\,,
\label{blackbody}
\end{equation}
where $A(M,n_i,d)$ is the BH area which is induced on the $n_i$-dimensional
subspace, and $c_i(n_i)$ and $\Gamma_i(E,n_i,s_i)$ are the number of degrees of
freedom and the greybody factor of the species $i$, respectively. Two remarks
are in order. First, we assume that the BH induced area depends only on $n_i$,
on the BH mass $M$, and on the number of spacetime dimensions $d$. In this
paper we are interested in spherically symmetric BHs, so this condition is
always verified. Second, Eq.~\refb{blackbody} is very well approximated by
considering the thermally averaged greybody factors \cite{Page:df}. Therefore,
we drop the $E$-dependence in the $\Gamma_i$'s and use the BH geometric optics
area \cite{Emparan:2000rs}:
\begin{equation}
A_i(M,n_i,d)=\Omega_{n_i-2}r_c^{n_i-2}\,,
\label{area}
\end{equation}
where $\Omega_{n_i-2}$ is the area of the unit $(n_i-2)$-dimensional sphere and
\begin{equation}
r_c=\left(\frac{d-1}{2}\right)^{\frac{1}{d-3}}
\left(\frac{d-1}{d-3}\right)^{1/2}r_s
\label{rc}
\end{equation}
is the optical radius of the $d$-dimensional Schwarzschild BH of radius $r_s$.
The emitted energy density distribution in $n_i$-dimensions, $d{\cal E}_{\rm
em}/dt$, is related to the blackbody energy  density distribution $d{\cal
E}_{{\rm BB},i}/dt=E\,d{\cal N}_i/dt$ by
\begin{equation}
\frac{d{\cal E}_{{\rm em},i}}{dt}=\frac{\Omega_{n_i-3}}{(n_i-2)\Omega_{n_i-2}}
\frac{d{\cal E}_{{\rm BB},i}}{dt}\,.
\label{energy}
\end{equation}
By integrating Eq.~\refb{energy} over the phase space and summing over all the
particle species we obtain the total emitted energy per unit time of a BH with
mass $M$ (generalized Stefan-Boltzmann equation):
\begin{equation}
\frac{dE_{\rm em}}{dt}=\sum_i \sigma_{n_i} 
c_i(n_i)\Gamma_i(n_i,s_i)f_i(n_i) A_i(M,n_i,d) T^{n_i}\,.
\label{sb}
\end{equation}
The $n_i$-dimensional Stefan-Boltzmann constant is
\begin{equation}
\sigma_{n_i}=\frac{\Omega_{n_i-3}\Gamma(n_i)\zeta(n_i)}{(n_i-2)(2\pi)^{n_i-1}}\,,
\label{sbconst}
\end{equation}
where $\Gamma$ is the gamma Euler's function (not to be confused with the
greybody factor $\Gamma_i$) and $f_i(n_i)=1$ $(1-2^{1-n_i})$ for bosons
(fermions). Using Eq.~\refb{area} and the relation between the BH temperature
and the Schwarzschild radius \cite{Argyres:1998qn},
\begin{equation}
T=\frac{d-3}{4\pi r_s}\,,
\label{temp}
\end{equation}
it follows that the energy emitted per unit time is proportional to the
temperature square:
\begin{equation}
\frac{dE_{\rm em}}{dt}=T^2\sum_i \sigma_{n_i} 
c_i(n_i)\Gamma_i(n_i,s_i)f_i(n_i)\mu_i(n_i,d)\,,
\label{sb2}
\end{equation}
where
\begin{equation}
\mu_i(n_i,d)=\Omega_{n_i-2}\left(\frac{d-1}{2}\right)^{\frac{n_i-2}{d-3}}
\left(\frac{d-1}{d-3}\right)^{\frac{n_i-2}{2}}
\left(\frac{d-3}{4\pi}\right)^{n_i-2}\,.
\label{mui}
\end{equation}
The ratio of the emitted energy for two different species is:
\begin{equation}
\frac{dE_{{\rm em},i}/dt}{dE_{{\rm em},j}/dt}=
\frac{\sigma_{n_i}c_i(n_i)\Gamma_i(n_i,s_i)f_i(n_i)\mu_i(n_i,d)}
{\sigma_{n_j}c_j(n_i)\Gamma_j(n_j,s_j)f_j(n_j)\mu_j(n_j,d)}\,.
\label{ratioe}
\end{equation}
Using Eq.~\refb{sb2}, the BH mass loss $dM/dt=-dE_{\rm em}/dt$ can be
expressed as a function of the BH mass:
\begin{equation}
\frac{dM}{dt}=-\mu(\{n_i,s_i\},d)M^{-\frac{2}{d-3}}\,.
\label{massloss}
\end{equation}
where
\begin{equation}
\mu(\{n_i,s_i\},d)=\left(\frac{d-3}{4\pi}\right)^2
\left[\frac{(d-2)\Omega_{d-2}}{16\pi}\right]^{\frac{2}{d-3}}
\sum_i \sigma_{n_i}c_i(n_i)\Gamma_i(n_i,s_i)f_i(n_i)\mu_i(n_i,d)\,.
\label{mu}
\end{equation}
Integrating Eq.~\refb{massloss} we obtain the decay time
\begin{equation}
\tau=\mu^{-1}\frac{d-3}{d-1}M_{\rm BH}^{\frac{d-1}{d-3}}\,.
\label{tau}
\end{equation}
We can compare this result to the decay time of Argyres {\it et al.}
\cite{Argyres:1998qn}, which is often used in the literature. The decay time of
Ref.~\cite{Argyres:1998qn} is calculated by taking into account only the
graviton loss in $d$-dimensions. Our result differs from the result of
Ref.~\cite{Argyres:1998qn} because the latter is calculated by using the actual
area of the BH rather than the optical area and neglecting the geometric
emission factor of Eq.~\refb{energy}. As is expected, the decay time
Eq.~\refb{tau} is longer than that of Ref.~\cite{Argyres:1998qn} because of the
extra particle evaporation channels.

The total multiplicity is obtained by integrating $E^{-1}\,{\cal E}_{{\rm
em},i}/dt$ over the phase space and summing over all the particle species. Using
Eq.~\refb{temp} and Eq.~\refb{massloss} we find
\begin{equation}
N=\frac{d-3}{d-2}\frac{M_{\rm BH}}{T_{\rm BH}}
\frac{\sum_i\sigma_{n_i}c_i(n_i)\Gamma_i(n_i,s_i)f_i(n_i-1)\mu_i(n_i,d)
\zeta(n_i-1)[(n_i-1)\zeta(n_i)]^{-1}}
{\sum_j\sigma_{n_j}c_j(n_j)\Gamma_j(n_j,s_j)f_j(n_j)\mu_j(n_j,d)}\,.
\label{multi}
\end{equation}
The multiplicity per particle species is
\begin{equation}
N_i=N\frac{\sigma_{n_i}c_i(n_i)\Gamma_i(n_i,s_i)f_i(n_i-1)\mu_i(n_i,d)
\zeta(n_i-1)[(n_i-1)\zeta(n_i)]^{-1}}
{\sum_j\sigma_{n_j}c_j(n_j)\Gamma_j(n_j,s_j)f_j(n_j-1)\mu_j(n_j,d)
\zeta(n_j-1)[(n_j-1)\zeta(n_j)]^{-1}}\,.
\label{multifla}
\end{equation}
Equation \refb{multifla} gives the statistical number of particles per
species produced during the evaporation process.

\section{Black hole evaporation at LHC}

Using the previous equations, we can estimate the energy loss, the decay
time, and the multiplicity of Schwarzschild BHs at particle colliders.
Equations \refb{ratioe} and \refb{tau}-\refb{multifla} are exact, modulo
the greybody thermal average approximation. These equations can be
evaluated for a given particle model (standard model, SUSY, etc\dots) and
a given geometry of the spacetime (ADD \cite{Arkani-Hamed:1998rs},
Randall-Sundrum \cite{Randall:1999ee, Randall:1999vf}, fat brane
\cite{Arkani-Hamed:1999dc}, universal extra dimensions
\cite{Appelquist:2000nn}, etc\dots) if the corresponding greybody factors
$\Gamma_i(n_i,s_i)$ are known. The greybody factors for particles with
spin 0, 1/2, 1 and 2 in four dimensions have been known for a long time
\cite{Page:df}. The greybody factors for fields with spin 0, 1/2, and 1 in
higher-dimensions have recently been calculated in the low-frequency limit
\cite{Kanti:2002nr, Kanti:2002ge, Ida:2002ez}. To our knowledge, the
spin-2 greybody factors in higher dimensions have not been calculated.
Therefore, some kind of approximation is required in Eq.~\refb{ratioe} and
Eqs.~\refb{tau}-\refb{multifla}.

Let us assume that the BH mainly radiates on the brane. Usually, only gravitons
and other ``exotic'' fields propagates in the whole $d$-dimensional spacetime.
Hence, one should expect the energy-loss into the bulk to be negligible. In
order to check this assumption, let us consider a simple model with a
four-dimensional infinitesimally thin brane and only the graviton propagating
into the 10-dimensional bulk. Using Eq.~\refb{ratioe} we can evaluate the ratio
of energy loss due to graviton emission in the bulk and, for example, the
total spin-1 field emission on the brane. Equation~\refb{ratioe} gives the
result:
\begin{equation}
\frac{dE_{\rm em,grav}/dt}{dE_{\rm em,spin\hbox{-}1}/dt}\approx .64\,
\frac{\Gamma_{\rm spin\hbox{-}2}(10,2)}{\Gamma_{\rm spin\hbox{-}1}(4,1)}\,,
\label{gravtospin1}
\end{equation}
where the spin-1 greybody factor is $\Gamma_{\rm spin\hbox{-}1}(4,1)\approx
0.24$. In four-dimensions the spin-2 greybody factor is $\Gamma_{\rm
spin\hbox{-}2}(4,2)\approx 0.028$, thus in a pure four-dimensional world the
graviton emission is negligible. However, if the ten-dimensional spin-2
greybody factor is bigger than $\approx .38$, {\em the graviton emission in the
bulk becomes comparable to the total spin-1 emission on the brane}! In this
case the bulk emission cannot be neglected. This is quite in contrast with the
result obtained by assuming a Boltzmann statistics and no greybody factors
\cite{Emparan:2000rs}. A definitive answer requires the knowledge of the
graviton greybody factors in higher dimensions.

We illustrate the computation of the multiplicity with an example. Let us
evaluate Eq.~\refb{multi} for a $n$-dimensional infinitesimally thin brane in a
$d$-dimensional spacetime. We consider only standard model fields and we
neglect the graviton energy loss into the bulk. Setting $n_i=n_j=n$,
Eq.~\refb{multi} simplifies to
\begin{equation}
N=\frac{d-3}{d-2}\frac{\zeta(n-1)}{(n-1)\zeta(n)}
\frac{M_{\rm BH}}{T_{\rm BH}}
\frac{\sum_i c_i(n)\Gamma_i(n,s_i)f_i(n-1)}
{\sum_j c_j(n)\Gamma_j(n,s_j)f_j(n)}\,.
\label{multisimpl}
\end{equation}
Setting $n=4$, we can evaluate the statistical multiplicity of BH events at LHC
as function of the BH mass and of the spacetime dimensions. Choosing the
fundamental Planck scale to be $M_{\rm Pl}=1$ TeV, the total number of emitted
particles is $N=4$, $6$, $7$ for $M_{\rm BH}=8$, $10$ and $12$ TeV,
respectively. Let us compare these values to the average multiplicity $\langle
N\rangle$ which is used in the literature. For the same choices of $M_{\rm BH}$
we find $\langle N\rangle=8$, $10$, and $12$. The multiplicity calculated by
taking into account the particle statistics and the greybody factors is reduced
by a factor $\approx 43$\% w.r.t.\ average multiplicity. The reduction is
stable as the number of spacetime dimensions changes, varying from $\approx
48$\% for $d=7$ to $\approx 42$\% for $d=11$. If we take into account possible
emission of gravitons in the bulk, the visible multiplicity could be further
reduced. How are the decay products distributed among the particle flavors?
Using Eq.~\refb{multifla} we find that the $M_{\rm BH}=8$ TeV BH is likely to
decay into three quarks plus either one charged lepton or one gluon, and the
$M_{\rm BH}=10$ ($12$) TeV BH decays into four (five) quarks, one charged
lepton and one gluon, each quark and gluon eventually producing a hadronic jet.
The hadron-to-lepton ratio is still about 5:1.

\section{Conclusion}

In this paper we have derived the energy loss, the statistical multiplicity and
the flavor multiplicity of spherically symmetric BHs by taking into account 
the Boltzmann statistics and the greybody factors of the decay products. We
have found that the multiplicity may be significantly smaller than the average
multiplicity which is  usually considered in the literature. The large
multiplicity has long be considered one of the main signatures of BH formation
at particle colliders. If the fundamental Planck scale is of order of the TeV
scale, we expect creation of BHs with mass of a few TeVs at LHC. The
multiplicity of these BHs is, however, far from being large, at least in the
simplest model of a brane with infinitesimal thickness. Assuming $M_{\rm Pl}=1$
TeV and ten dimensions, BHs in the range $8$ - $10$ TeV produce at most four or
five hadronic jets. This has important consequences on the possible detection
of BHs events, making the discrimination between standard model events and
gravitational events harder. The introduction of the greybody factors and of
the particle statistics also affects the bulk-to-brane energy loss ratio.
Usually, the bulk emission is considered negligible w.r.t.\ emission into the
brane. However, the exact amount of energy radiated into the bulk vs.\ the
brane crucially depends on the greybody factors even in the simplest models.
Since no greybody factors for spin-2 field in higher dimensions have been
calculated, the answer to the question whether the emission in the bulk is
negligible remains open. If the BH energy loss into the bulk is comparable with
the energy loss on the brane, the possibility of BH detection in particle
colliders is further reduced.  

\section*{Acknowledgments}
We thank E.-J.~Ahn, S.~Das, R.~Maartens for interesting discussions and
comments. This work is supported by PPARC.

\end{document}